\documentstyle[12pt,aaspp4]{article}

\received{}
\revised{}
\accepted{}
\journalid{}{}
\articleid{}{}
\paperid{}
\cpright{}{}
\ccc{}

\slugcomment{Submitted to the Astrophysical Journal}

\lefthead{Wood \& Raymond}
\righthead{Optically Thick Resonance Line Scattering}

\begin{document}

\title{Resonant Scattering of Emission Lines in Coronal Loops: 
Effects on Image Morphology and Line Ratios}

\author{Kenneth Wood and John Raymond}

\affil{Harvard-Smithsonian Center for Astrophysics,
60 Garden Street, Cambridge, MA 02138\\  
kenny@claymore.harvard.edu, jraymond@cfa.harvard.edu}

\authoremail{kenny@claymore.harvard.edu}

\begin{abstract}

We have investigated the effects of resonant scattering of emission lines
on the image morphology and intensity from coronal loop structures.  
It has previously been shown that line of sight effects 
in optically thin line emission can yield loop images that appear uniformly 
bright at one viewing angle, but show ``looptop sources'' at other viewing 
angles.  For optically thick loops where multiple resonant scattering is 
important, we use a 3D Monte Carlo radiation transfer code.  Our simulations 
show that 
the intensity variation across the image is more uniform than the optically 
thin simulation and, depending 
on viewing angle, the intensity may be lower or higher than that 
predicted from optically thin simulations due to scattering out of or into 
the line of sight.

\end{abstract}


\section{Introduction}

Two effects of resonant line scattering on emission lines are 
to change the morphology of images and to change line ratios from what 
would be expected from optically thin analyses.  Qualitatively, the 
effects can be split into three broad optical depth categories, low, 
moderate, and large.  At low optical depths ($\tau < 5$) resonant scattering 
primarily changes the photon direction and photons emerge into the 
directions of lowest optical depth.  At moderate optical depths 
($\tau \sim 10 - 100 $) multiple scattering is important and there is a large 
probability of photon energy conversion, e.g., the conversion of Ly$\beta$ 
photons to H$\alpha$ photons.  Large optical depths ($\tau >> 100$) 
result in photon trapping and collisional de-excitation, eventually leading
to thermalization.  The goal of this 
paper is to investigate the effects of resonant scattering on images and 
intensity ratios in the low optical depth regime. 
Situations where our work is applicable 
include UV, EUV, and soft X-ray emission line imaging and spectroscopy of 
the Sun and stellar coronae.  While we have chosen a geometry based on
coronal magnetic loops, similar considerations would apply, for instance,
to some of the strong X-ray emission lines in older supernova remnants.
We have simulated an X-ray line of Fe~{\sc xvii},
but the predicted effects on image morphology should be visible in
the narrow-band EUV images of EIT and TRACE.

Ultraviolet and soft X-ray emission line ratios are frequently used as 
diagnostics of temperature, density, elemental abundances, and the ionization 
state of astrophysical plasmas.  Several assumptions are often implicitly 
made when emission line ratios are analyzed.  In particular, the optical 
depth is generally assumed to be negligible.  
However, solar coronal 
observations of the Fe~{\sc xvii} line at 15.01{\AA} show line ratios 
that are lower than theoretical predictions (Phillips et al. 1997; 
Schmelz et al. 1997; Saba et al. 1999; Waljeski et al. 1994) or 
laboratory measurements (Brown et al. 1998).  
Rugge \& MacKenzie (1985)
suggested that optically thick resonant scattering of 
Fe~{\sc xvii} 15.01{\AA} photons out of the line of sight may be responsible 
for the small line ratios.  
Saba et al. (1999) employed the escape probability method of 
Kastner \& Kastner (1990) in their analysis of various line ratios and 
found that up to 50\% of the Fe~{\sc xvii} flux is unaccounted for.  
Bhatia \& Kastner (1999) found similar results for planar and cylindrical 
geometries.  
This explanation for the intensity 
decrements is likely, given the large oscillator 
strength of the Fe~{\sc xvii} 15.01{\AA} line (Table~1).  
For typical densities 
($n_e \sim 10^{10}$) and sizescales ($\sim10^9$cm) in the solar corona, the 
optical depth of the Fe~{\sc xvii} 15.01{\AA} line is around 1.  
This is in the 
low optical depth regime in which the major effect is scattering of photons 
into optically thin directions.

In this paper we present three dimensional Monte Carlo radiation transfer 
simulations of the transfer of Fe~{\sc xvii} 15.01{\AA} photons.  
Our analysis does not treat transfer 
within the line itself, instead considering all photons to be emitted and 
resonantly scattered at a single wavelength.  This approximation is valid so 
long as the optical depth is not large enough that photons scatter out
of the Doppler core.  
The advantage of our Monte Carlo technique is that it naturally 
accounts for arbitrary illumination and multiple scattering in complex 
geometries.  The results of our Monte Carlo simulations are images that may be 
compared directly with optically thin simulations to determine differences 
in image morphology and optically thick to optically thin intensity ratios.

Previous simulations by Alexander \& Katsev (1996) produced optically thin 
images of loops in which the density and temperature follow the scaling law 
prescription of Rosner, Tucker, \& Vaiana (1978, hereafter RTV).  
The simulations of 
Alexander \& Katsev showed that it is crucial to consider the geometry 
and temperature dependence of the emissivity along with the thermal 
response of the imaging telescope when analyzing solar emission line 
images.  Such images may be in the light of a single line or an ensemble 
of lines as is the case with the broad wavelength response of Yohkoh SXT.  
In particular, a loop viewed face-on may appear uniformly bright, but 
line-of-sight effects can yield an apparent ``looptop'' source when the same 
loop is viewed close to edge-on (see Fig.~5 of Alexander \& Katsev).  Our work 
is a natural extension of the Alexander \& Katsev analysis to optically thick, 
multiple scattering situations, and for continuity purposes we also adopt 
RTV loops for our simulations.

Section~2 presents the model ingredients (loop structure, Fe~{\sc xvii} 
emissivity and opacity, and radiation transfer technique).  Results of our 
simulations (images and intensity ratios 
of optically thick/optically thin models) are presented in \S~3, and 
in \S~4 we discuss our results with respect to observational data.

\section{Model Ingredients}

In order to construct model images we must specify the emissivity, 
opacity, and scattering geometry within our Monte Carlo radiation 
transfer code.  These model ingredients are now described.

\subsection{Loop Structure}

Following Alexander \& Katsev, we consider loops with a density and 
temperature structure given by the RTV scaling law model.  The RTV model is 
a hydrostatic loop model in which the maximum loop temperature, 
$T_{\rm max}$, and loop length, $L$, are related to the pressure, $p$, by 
$T_{\rm max}\sim 1.4\times10^3(pL)^{1/3}$.  The maximum temperature occurs 
at the loop apex, and the balance among thermal conduction, radiative cooling
and mechanical heating determines the temperature distribution
along the length of the loop.  An RTV loop is therefore 
specified by two parameters, the maximum temperature, $T_{\rm max}$, and 
loop length, $L$, measured along the loop from a footpoint to the looptop.  
Throughout our simulations we assume the loops are semicircular tori with 
length $L=10^{10}$cm and radius $r=L/10$.  We then construct simulations 
for a range of maximum temperatures and viewing angles.  See Fig.~1 for 
a diagram showing the loop length, radius, and orientation.

In some cases we embed the emitting loop in a lower density background plasma.
Alexander \& Katsev modeled isolated loop structures.  We wish to 
investigate situations where a background plasma may provide 
additional scattering opacity, thereby allowing for attenuation 
of loop emission and the formation of scattering halos around the loops.  
For the background we embed the RTV loop in a hemispherical isothermal 
constant density plasma.  We adopt a temperature of 
$T=2.5\times 10^6$K and set the density such that the total emissivity 
from the hemispherical volume is half of that emanating from the RTV loop 
(see \S~3).

We note here that an isothermal, constant density background 
plasma may be physically unrealistic.  
The precise form of the background is unimportant for the 
purposes of this work, as we wish to show the general effects of resonant 
scattering.  For future modeling of images from TRACE, EIT, and SXT 
where resonant scattering effects are important, more realistic loop 
structures and backgrounds will be generated.

\subsection{Fe~{\sc xvii} Emissivity and Opacity}

The Fe~{\sc xvii} 15.01 {\AA} line 
emissivity and opacity are functions of temperature within 
the loop.  The Fe abundance was taken as the typical FIP (first ionization 
potential) enhanced value of three times the photospheric abundance 
($1.2 \times 10^{-4}$; e.g. Meyer 1985; Feldman 1992).
We have taken the Fe~{\sc xvii} ionic fraction as a function of temperature 
from the paper of Arnaud \& Raymond (1992), which is similar to the 
more recent 
results of Mazzotta et al. (1998).  Because of its stable Ne-like electron 
configuration, Fe~{\sc xvii} is the dominant ion over a broad temperature 
range, and its emission lines are among the brightest observed in 
astrophysical plasmas at temperatures around 
$3\times 10^6$ K.  The excitation rate for the 15.01 {\AA} line 
($2p^6 - 2p^5 3d\; ^1P_1$, Table~1)
is taken from Smith et al. (1985), and it is very close to the results 
of other computations (e.g., Goldstein et al. 1989).  The oscillator strength 
is from Bhatia \& Doschek (1992).  To obtain optical depths, we assume a 
line width of 
$30{~\rm km~s^{-1}}$, typical of the solar corona (Mariska, 
Feldman \& Doschek 1979).
The emissivity and opacity for the range of temperatures we consider are 
displayed in Fig.~2.  

\subsection{Radiation Transfer}

We perform the radiation transfer with a Monte Carlo radiation transfer 
code that accounts for arbitrary sources of emission and multiple scattering 
within an arbitrary geometry.  The scattering code is based on that described 
by Code \& Whitney (1995), but has now been modified to run on a three 
dimensional linear Cartesian grid (Wood \& Reynolds 1999) and includes  
forced first scattering (Witt 1977) and a ``peeling off'' procedure 
(Yusef-Zadeh, Morris, \& White 1984).  These modifications enable us to 
construct model images of three dimensional systems from specified viewing 
angles very efficiently.

For the current investigation we set up our grid with an RTV loop structure.  
The temperature dependent emissivity and opacity within the loop are 
determined as described above.  We assume that all Fe~{\sc xvii} photons 
are emitted and resonantly scattered (with an isotropic scattering phase 
function) at a single wavelength.  It is possible that when a 15.01 \AA\/
photon is absorbed, the excited level will decay to a 3p level, converting
the energy into two EUV photons and a 3s-2p photon near 17 \AA .  However,
the branching ratio for this conversion is only $5.7 \times 10^{-4}$,
which can safely be ignored for the optical depths less than about 10 that
we consider.  We also assume that the photon does not leave the Doppler core.
This is a good approximation for modest optical depths, though it would seriously
underestimate the flux emerging in the line wings for large optical depths.
We use the average scattering cross section over a Gaussian profile, 
rather than using the actual cross section as a function of velocity for 
computational speed.

\section{Results}

For each simulation of an RTV loop ($L=10^{10}$cm) 
we form images at different viewing angles.  
In \S~3.1 we show some representative Monte Carlo images 
that illustrate the effects of optical depth and viewing angle.
From these images we compute the total emergent intensity for the 
particular viewing 
angle for comparison  with an optically thin simulation (\S~3.2).

\subsection{Images}

In Figs.~3, 4, and 5 we show a set of model images for loops of different 
$T_{\rm max}$ at different viewing angles.  We show optically thin images 
(assuming zero optical depth), 
and optically thick images with and without background plasma.  In this and 
the following section we refer to viewing angles $\theta$ and $\phi$, where 
the polar angle, $\theta$, is measured from the $z$ axis and the 
azimuthal angle, $\phi$, is measured anti-clockwise from the $x$ axis.  
In a Cartesian coordinate system, the loop is oriented in the $z-y$ plane 
(see Fig.~1).  
There are several features of the images that we now describe.

The $T_{\rm max}$ dependence affects the distribution of the emissivity along 
the loop.  For the lowest temperature, $\log T_{\rm max} = 6.4$ (Fig.~3), the 
emissivity is concentrated towards the loop apex.  This is because at 
$\log T_{\rm max} = 6.4$ the Fe~{\sc xvii} 15.01{\AA} line emissivity is 
almost zero (Fig.~2).  This temperature occurs at the loop apex and 
the loop legs are cooler resulting in the concentration of emission at the 
apex and negligible emission from the loop legs.  The emissivity distribution 
along the loop changes as we increase $T_{\rm max}$.  
As $T_{\rm max}$ increases ($\log T_{\rm max} = 6.6$ in Fig.~4 and 
$\log T_{\rm max} = 6.8$ in Fig.~5) the emissivity becomes more concentrated 
towards the legs and footpoints of the loop.  For $\log T_{\rm max} = 6.8$, 
the Fe~{\sc xvii} 15.01{\AA} line emissivity is again almost zero (Fig.~2) 
and in this case the 
emission is concentrated towards the loop legs and there is negligible 
emission from the loop apex.

The geometrical line-of-sight effects for the optically thin images 
discussed by Alexander \& Katsev are clearly seen in column~A 
of Figs.~3, 4, and 5.  When viewed from above, geometrical projection makes 
the footpoints appear bright (image A1 in Figs.~4 and 5), while the 
apex of the loop is brightened by projection if the loop is viewed edge-on 
(image A5 in Figs.~4 and 5).  
Loops that appear uniformly bright when seen face-on (image A3 in 
Figs.~3 and 4)
appear to have bright looptops when viewed edge-on (image A5 Figs.~3 and 4).  
Similarly, loops that show brightening along the legs for face-on viewing 
(image A3 in Fig.~5),
appear more uniformly illuminated when viewed edge-on (image A5 in Fig.~5).  
This is 
simply due to the increased path length towards the apex for edge-on loops.

Optical depth drastically reduces the geometrical projection effects 
because the long path length that enhances the brightness of an optically 
thin line implies a large optical depth for an optically thick line.  
The effect on image morphology is that isolated loops appear 
more uniformly bright (compare columns~A and B in Figs.~3, 4,
and 5).  This arises because 
we see emission originating from the ``optical depth one surface'' which 
covers a progressively larger projected area on the loop as it becomes 
optically thicker.  This effect is shown quantitatively in Fig.~6 where we 
show the intensity variation across two of the loop images from Fig.~4.  
Figure~6a shows the intensity variation 
across the images in Fig.~4 for the overhead 
view (images A1 and B1 in Fig.~4).  The solid line shows the 
intensity variation across the optically thin image (image A1 in Fig.~4) 
with the intensity peak towards the legs of the loop evident.  The 
dotted line shows the intensity variation for the same loop, but with the 
effects of resonant scattering included (image B1 in Fig.~4) --- the 
overall intensity level is lower 
and the intensity variation is more uniform.  Figure~6b shows the intensity 
variation across the images in Fig.~4 for the edge-on view 
(images A5 and B5 in Fig.~4).  
Again the solid line shows the intensity variation along 
the optically thin image (image A5 in Fig.~4) showing the intensity 
peak at the loop apex.  The dotted line shows a much reduced 
intensity level and 
more uniform intensity for the loop model that includes resonant scattering 
effects (image B5 in Fig.~4).

Real solar active regions are complex structures, with loops of differing
$T_{\rm max}$, length and pressure.  The emissivity scales as $n_e^2$, while
the opacity scales as $n_e$, so that dense, low-lying loops may dominate
the emission, but a background plasma may provide additional 
Fe~{\sc xvii} scattering opacity.  We simulate this effect by embedding
the emitting loop in a lower density background (column~C in 
Figs.~3, 4, and 5).  
The total emission from the background is half that from the RTV loop 
but the background is very diffuse (see Table~2), so the RTV 
loops are clearly visible.  

Two main effects of the background 
opacity are easily seen in Fig.~5.  The first is the scattered 
light halo present around the loops (e.g., image C5 in Fig.~5), 
arising from scattering of loop emission 
in the background plasma.  Such halos might appear as fuzziness in EIT or
TRACE images in EUV emission lines.  
The second effect is that the loop emission must 
pass through the background plasma and when the optical depth is large this 
results in scattering out of the line of sight and an asymmetry 
between the near and far side of the loops, resulting in an apparent 
one-sided loop (e.g., image C4 in Fig.~5).  
Asymmetric loops are often observed in SOHO/EIT and TRACE 
images (examples may be seen on the TRACE homepage: 
http://vestige.lmsal.com/TRACE/Public/Gallery/Images/).  
Most are probably asymmetric in reality, but optical depth may play 
a role in some observed morphologies.  

We note that the 171 and 195 \AA\/ bands of EIT and TRACE and 
the 284 \AA\/ band
of EIT are dominated by resonance lines of Fe~{\sc ix}, Fe~{\sc x}, 
Fe~{\sc xii} and Fe~{\sc xv}.
While these ions do not occupy as large a temperature range as Fe~{\sc xvii},
the scattering cross section scales in proportion to wavelength, and images
in these lines are likely to show similar effects to those predicted for
Fe~{\sc xvii} (it is the line optical depth and not the wavelength that 
determines the image morphology in our simulations).  
Indeed, Schrijver \& McMullen (2000) have inferred that scattering
effects are important from the variation of intensity with height above
the limb from TRACE and EIT images.

\subsection{Unresolved Intensity}

In the previous section we showed spatially resolved model images in the 
light of the Fe~{\sc xvii} line, as may be observed in the solar corona.  
Observations of stellar coronae cannot spatially resolve coronal loops and 
the CHANDRA satellite has recently observed Capella and other active 
late-type stars as part of the Emission Line Project.  As our resolved 
images showed, optically thick resonant scattering yields image 
morphologies and intensity distributions along the loops that differ from 
optically thin simulations.  In this section we investigate the effects 
of optically thick resonant scattering on the Fe~{\sc xvii} line 
intensity for different orientations of unresolved loops, 
and compare it with the intensity 
expected if the line were optically thin.  
For optically thin emission all emitted photons escape from the source 
isotropically, and all loops will yield the same intensity 
independent of viewing angle.  Optical depth effects result in photons 
exiting the simulation region preferentially along optically thinner 
sightlines.  Also, for optically thick loops, the intensity is proportional 
to the projected area of the loop, so face-on loops will yield larger 
intensities than edge-on loops.

Figure~7 shows the intensity ratio for an unresolved optically thick model 
compared to the corresponding model when optical depth effects are ignored.  
Each panel shows the intensity ratio for a range of $T_{\rm max}$, with the 
individual curves showing different viewing angles of the loop.  

The left column shows intensity 
ratios for an isolated RTV loop (no background 
plasma).  Face-on viewing ($\phi=0$, $\theta=\pi/2$) results in intensities 
that are larger than the optically thin case, due to photons being
scattered into our line of sight.  For edge-on viewing 
($\phi=\pi/2$, $\theta=\pi/2$), the intensity is lower than the 
optically thin case.  This is for two reasons; firstly the projected emitting 
area is smaller for edge-on loops (see the loop images in Figs.~3, 4, 
and 5).  Secondly as the optical depth increases, photons from the far side 
of the loop have to traverse a greater optical depth through the loop and are 
more likely to be scattered into the optically thinner face-on directions.  
Depending on $T_{\rm max}$ and loop orientation, the intensity can range
from being 20\% brighter to 50\% fainter than the intensity from an 
optically thin loop.

The second column in Fig.~7 shows intensity ratios for loops embedded within 
a background plasma.  The background has a temperature of 
$T_{\rm back}=2.5\times 10^6$K and the density is chosen so that the 
total intensity from the background is half that from the loop (Table~2).  
The optically thick to optically thin intensity 
ratios for these simulations show 
that the intensity ranges from around 60\% to 120\% of the optically thin 
intensity depending on $T_{\rm max}$ and viewing angle.  
These simulations show that 
optical depth, geometry, and loop orientation can have a large effect on 
observed line ratios.  Clearly great care is required in interpreting coronal 
images and line ratios due to the effects of projection and optically thick 
resonant scattering.

\section{Summary}

Our simulations have shown that optical depth effects, both within loops and 
in a background plasma, can change the amount of intensity emerging 
into a given direction 
and also the image morphology.  Intensity ratios may vary in the 
range 50\% to more 
than 120\% of their optically thin values.  If one observes an ensemble of 
loops, for instance spread over the surface of another star, the effects tend 
to average out.  Equal numbers of photons are scattered into and out of the 
line of sight.  However, a net effect may still arise if the emissivity and 
opacity vary in different ways.  Given the electron density factor in the 
emissivity and its absence in the opacity, such differences are likely.  
To account for the general trend of reduced Fe~{\sc xvii} 15.01 \AA\/ 
intensity
(e.g. Saba et al. 1999) by scattering rather than errors in atomic rates,
the typical structure must preferentially scatter photons down to the 
solar surface.  Such an example may be  emission from dense, low-lying 
loops that is 
scattered in a more diffuse, larger scale region.  This is expected from 
RTV models and is observed in solar active regions.  
Models similar to those in the 
righthand column of Fig.~5 can easily be constructed to reduce the intensities 
of lines having large oscillator strengths.  They may be invoked 
when line ratio anomalies appear in Chandra or XMM data.  Proving that optical 
depth is indeed the explanation for a given anomaly will undoubtably be a 
challenge.

\acknowledgements

KW acknowledges support from NASA's Long Term Space Astrophysics Research 
Program (NAG5-6039), and JR acknowledges support from NASA grant 
NAG5-2845.  We thank David Alexander, Carolus Schrijver, Rebecca McMullen, 
Julia Saba, and Danuta Dobrzycka for comments on this paper.  We also 
acknowledge the referee, K.R.~Phillips, for suggestions that improved the 
clarity of this paper.

\begin{deluxetable}{lcc}
\tablenum{1}
\tablewidth{0pt}
\tablecaption{Oscillator Strengths for Various Fe~{\sc xvii} Lines}
\tablehead{ \colhead{$\lambda$ ({\AA})} & \colhead{Transition} & \colhead{f} }
\startdata
15.01 & $2p^6 - 2p^5 3d\; ^1P_1$ & 2.66  \\
15.25 & $2p^6 - 2p^5 3d\; ^3D_1$ & 0.59  \\
15.46 & $2p^6 - 2p^5 3d\; ^3P_1$ & 0.0089  \\
16.79 & $2p^6 - 2p^5 3s\; ^1P_1$ & 0.10  \\
17.05 & $2p^6 - 2p^5 3s\; ^3P_1$ & 0.12  \\
17.10 & $2p^6 - 2p^5 3s\; ^3P_2$ & 0.00  \\
\enddata
\end{deluxetable}

\begin{deluxetable}{lcc}
\tablenum{2}
\tablewidth{0pt}
\tablecaption{RTV Loop $T_{\rm max}$, Background Temperature and Density}
\tablehead{\colhead{$\log {T_{\rm max}}$} &\colhead{$\log {T_{\rm back}}$}  
& \colhead{$n_{\rm back}$ (cm$^{-3}$)} }
\startdata
6.4 & 6.4 & 0.24$\times 10^{8}$ \\
6.5 & 6.4 & 0.70$\times 10^{8}$ \\
6.6 & 6.4 & 1.60$\times 10^{8}$ \\
6.7 & 6.4 & 3.00$\times 10^{8}$ \\
6.8 & 6.4 & 6.00$\times 10^{8}$ \\
\enddata
\end{deluxetable}

\begin{figure}[t]
\centerline{\plotfiddle{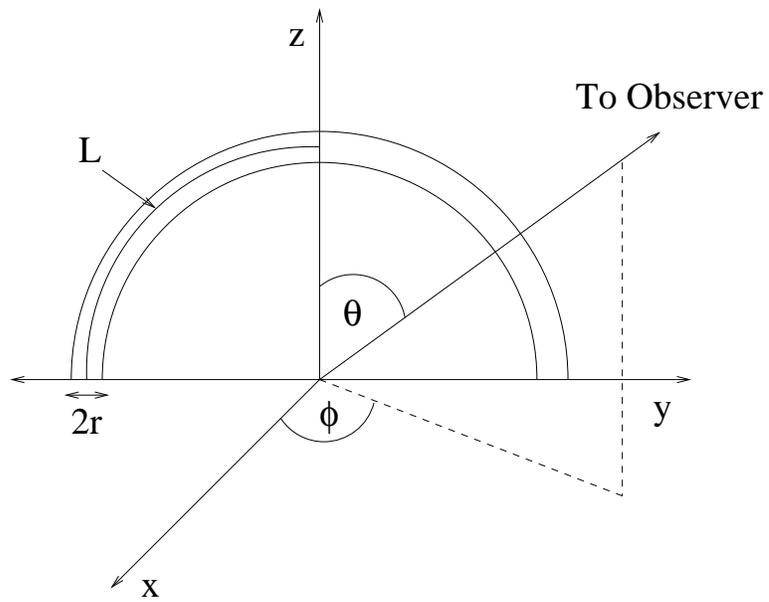}{6in}{0}{65}{65}{-420}{-20}}
\caption{Diagram showing the length, $L$, radius, $r$, and orientation 
($\theta$, $\phi$) of an RTV loop in our simulations.}
\end{figure}

\begin{figure}[t]
\centerline{\plotfiddle{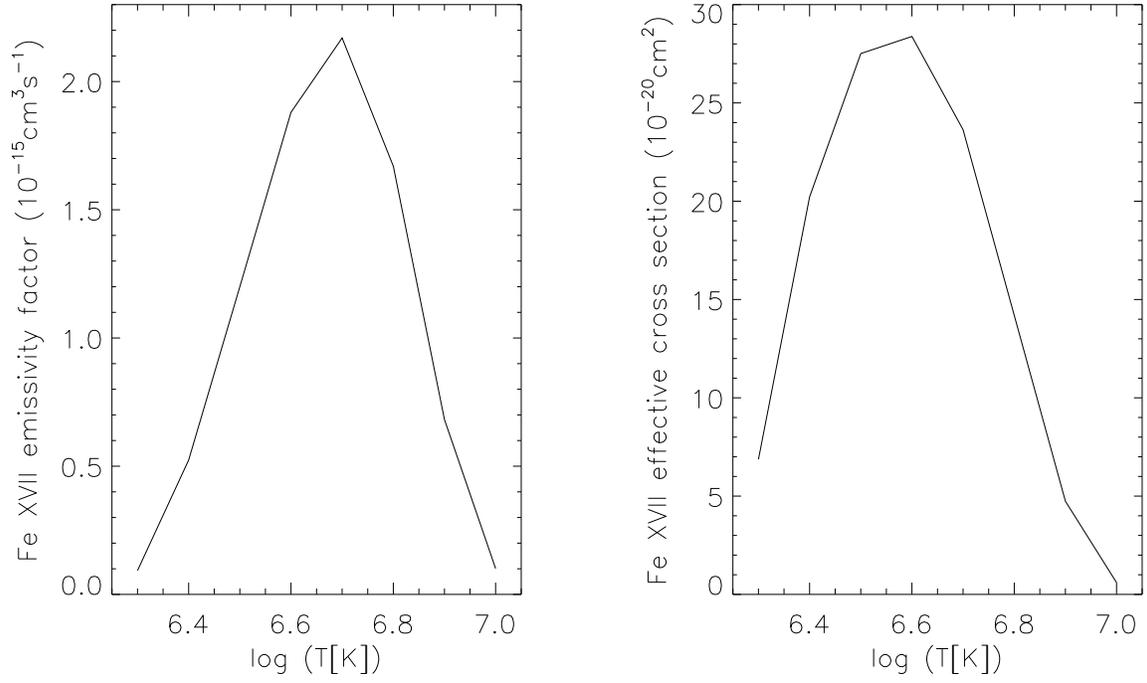}{6in}{90}{65}{65}{20}{100}}
\caption{Emissivity and effective scattering cross section for the 
Fe~{\sc xvii} 15.01{\AA} line as a function of temperature.  The effective 
cross section includes the Fe abundance and Fe~{\sc xvii} fractional 
concentration.}
\end{figure}

\begin{figure}[t]
\centerline{\plotfiddle{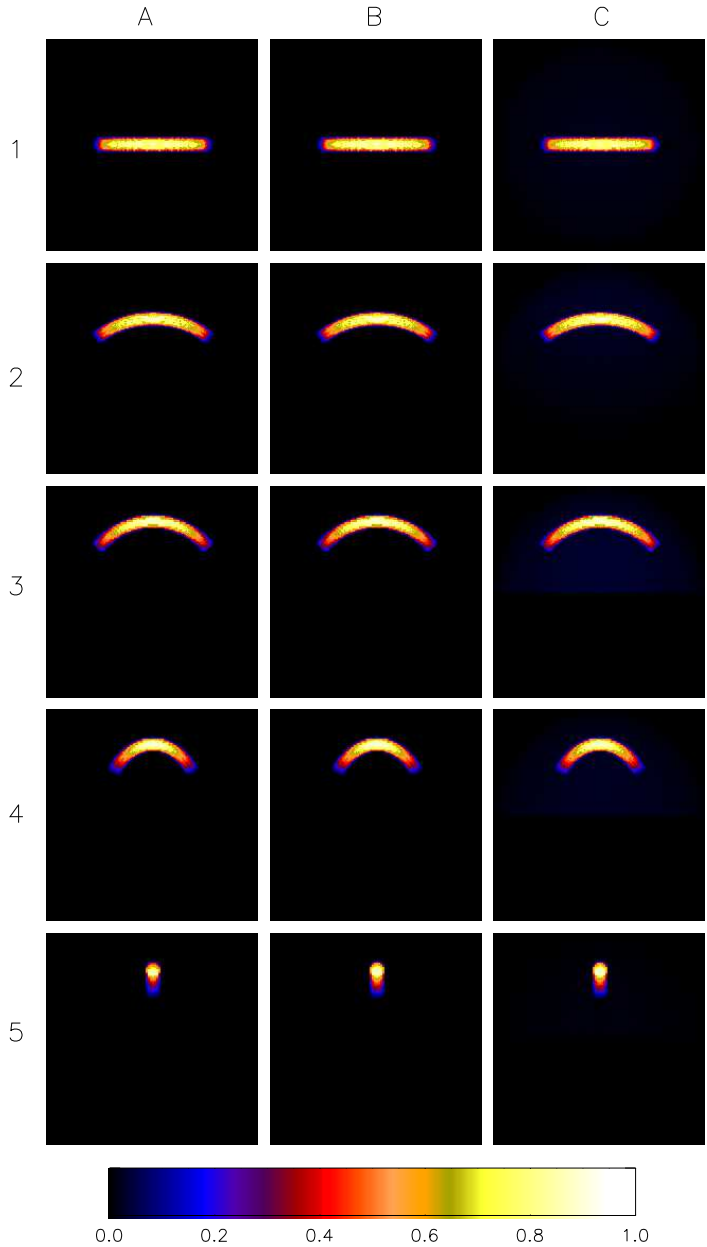}{6in}{0}{65}{65}{-420}{-20}}
\caption{Loop images at different viewing angles for 
$\log T_{\rm max} = 6.4$, where $T_{\rm max}$ is the maximum temperature of 
the RTV loop.  Each column shows appearance of a loop as viewed 
from (top to bottom) directly above (1), $45^\circ$ from vertical (2), 
face-on (3), rotated by $45^\circ$ (4), and edge-on (5).  
Column~A shows optically thin 
images, column~B is an optically thick isolated loop, column~C 
is a loop embedded in a background plasma 
($T=2.5\times 10^6$K, $n_e=2.4\times 10^7$cm$^{-3}$).  
Each image is $3\times 10^{10}$cm on a side.}
\end{figure}

\begin{figure}[t]
\centerline{\plotfiddle{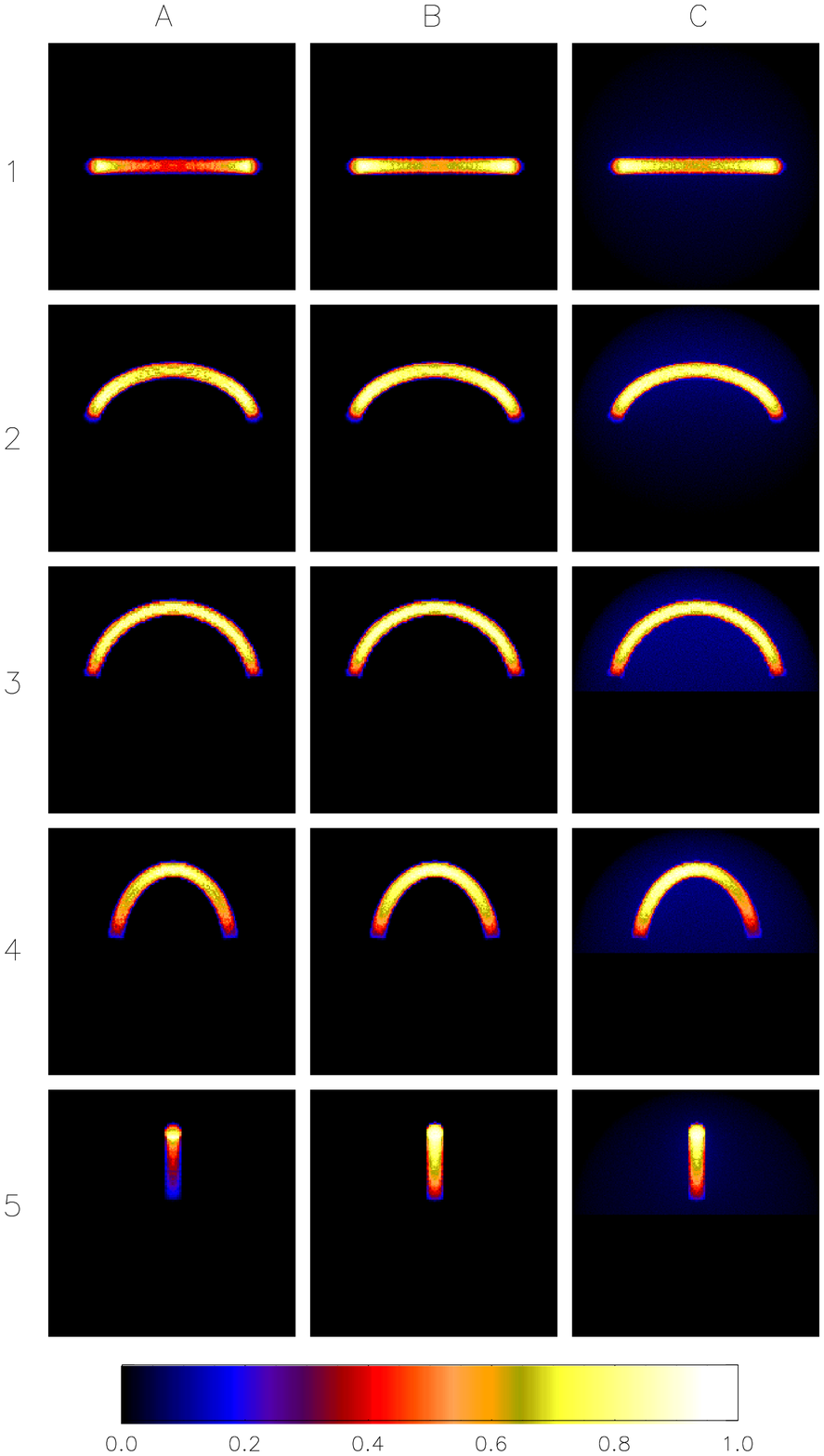}{6in}{0}{65}{65}{-420}{-20}}
\caption{As Fig.~3, but $\log T_{\rm max} = 6.6$ and a background plasma 
for column C of $T=2.5\times 10^6$K, $n_e=1.6\times 10^8$cm$^{-3}$.}
\end{figure}

\begin{figure}[t]
\centerline{\plotfiddle{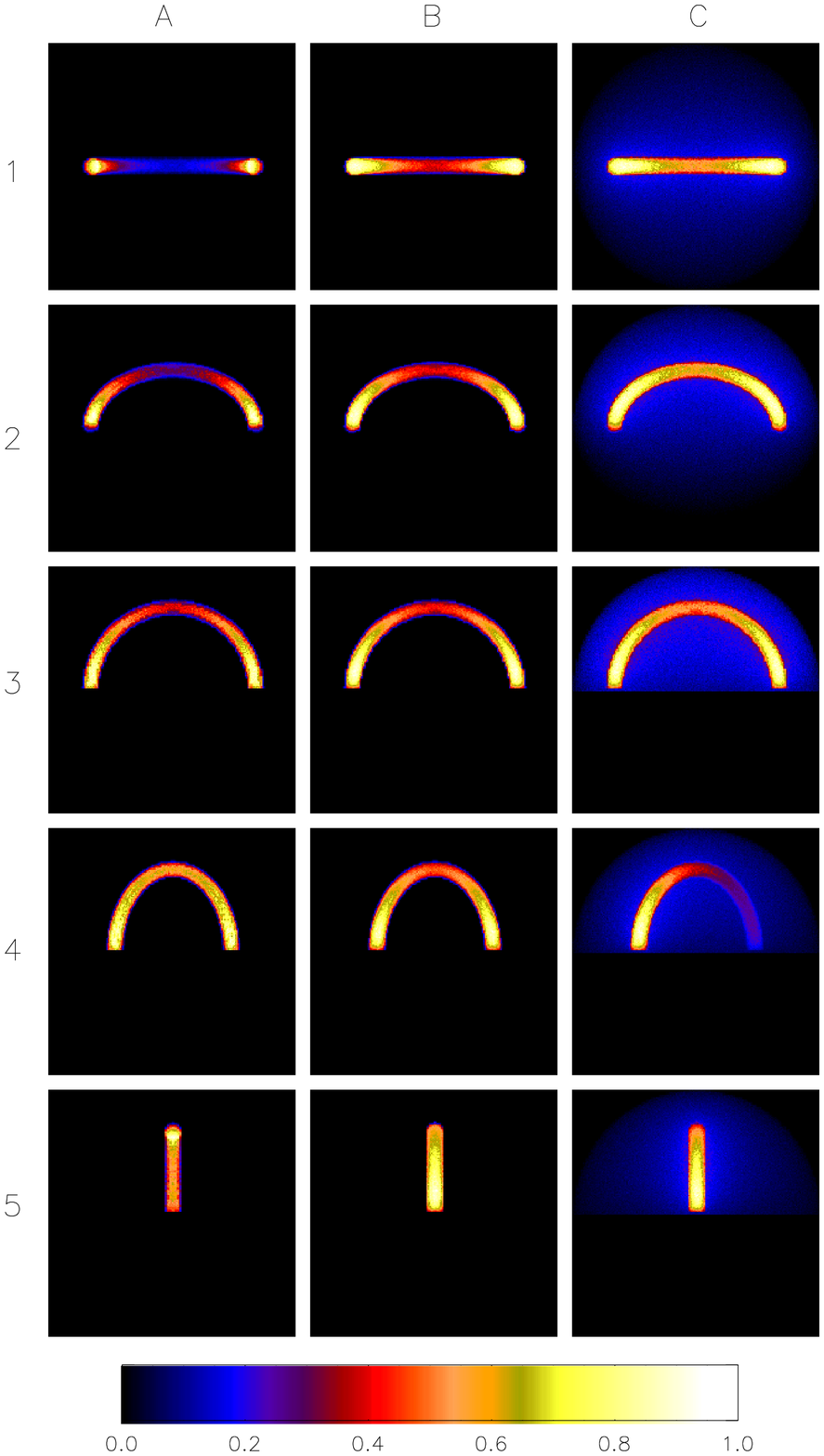}{6in}{0}{65}{65}{-420}{-20}}
\caption{As Fig.~3, but $\log T_{\rm max} = 6.8$ and a background plasma 
for column C of $T=2.5\times 10^6$K, $n_e=6\times 10^8$cm$^{-3}$.}
\end{figure}

\begin{figure}[t]
\centerline{\plotfiddle{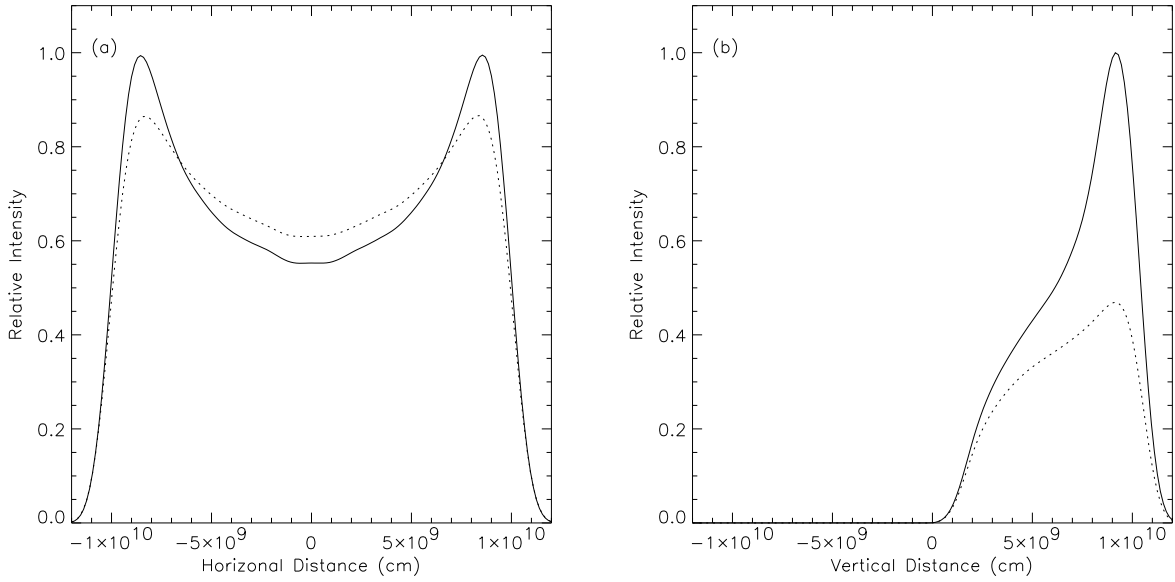}{6in}{90}{65}{65}{20}{60}}
\caption{(a) Solid line: intensity variation across image A1 in 
Fig.~4 (loop viewed from above) 
showing the intensity peak towards the loop legs for this optically 
thin simulation.  Dotted line shows intensity variation for the same loop 
model, but including resonant scattering effects (image B1 in Fig.~4).  
(b) As for (a), but for images A5 and B5 in Fig.~4 (loop viewed edge-on).}
\end{figure}

\begin{figure}[t]
\centerline{\plotfiddle{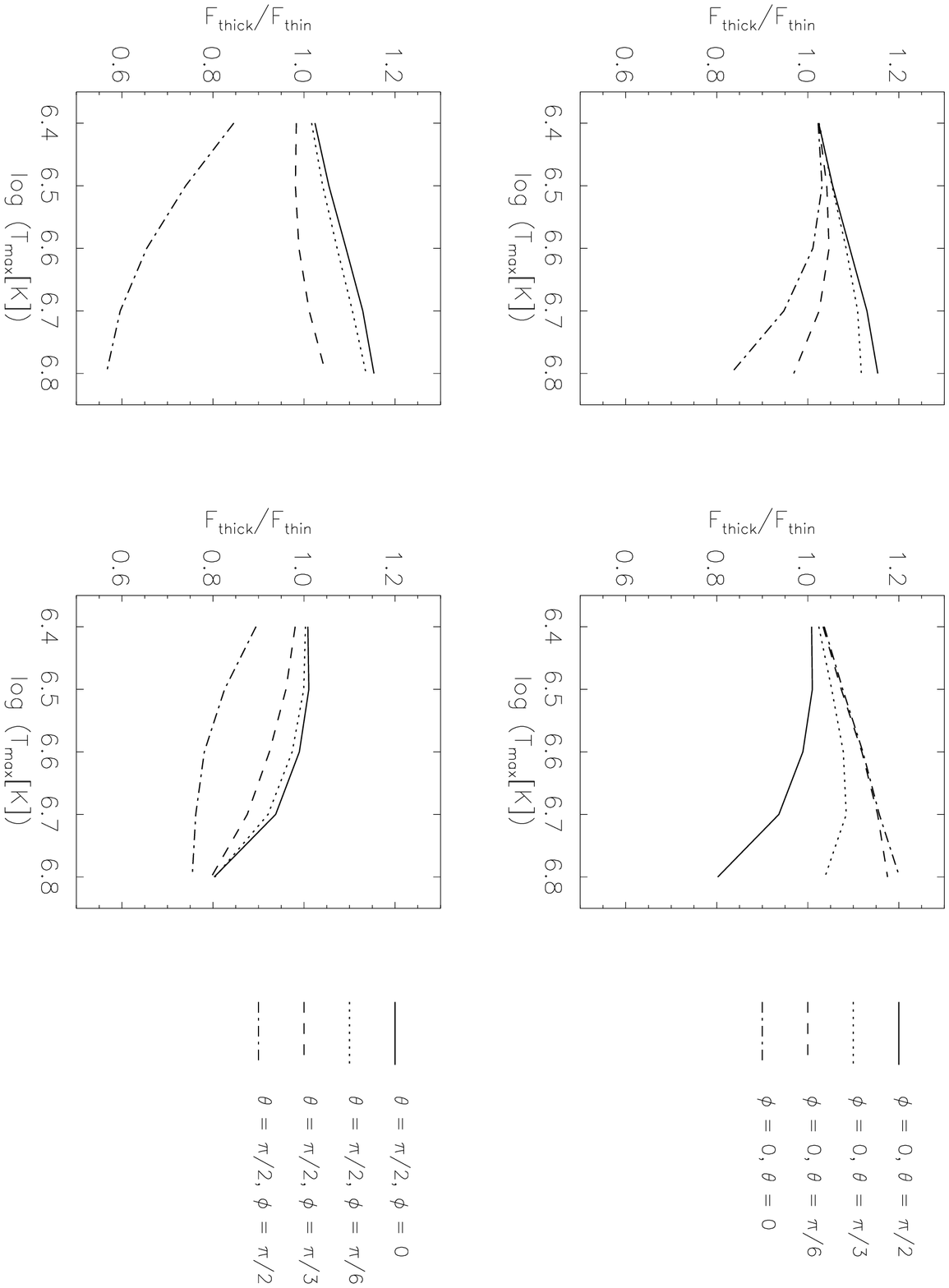}{6in}{90}{65}{65}{20}{60}}
\caption{Optically thick to optically thin intensity ratios as a function of 
maximum loop temperature.  Each panel shows the intensity ratios for different 
loop viewing angles.  Left most column shows ratios for isolated loops, 
second column for a loop embedded in a background plasma.  
Line types are for different viewing angles, 
upper row rotating through $\theta$, lower row rotating through $\phi$.}
\end{figure}


\begin{thebibliography}{}

\bibitem[]{}
Alexander, D., \& Katsev, S. 1996, Sol. Phys. 167, 153

\bibitem[]{}
Arnaud, M.A., \& Raymond, J.C. 1992, ApJ, 398, 394

\bibitem[]{}
Bhatia, A.K., \& Kastner, S.O. 1999, ApJ, 516, 482

\bibitem[]{}
Bhatia, A.K., \& Doschek, G.A. 1992, ADNDT, 52, 1

\bibitem[]{}
Brown, G.V., Beiersdorfer, P., Leidahl, D.A., Widmann, K., \& Kahn, S.M. 
1998, ApJ, 502, 1015

\bibitem[]{}
Code, A.D., \& Whitney, B.A. 1995, ApJ, 441, 400

\bibitem[]{}
Feldman, U. 1992, Physica Scripta, 46, 202

\bibitem[]{}
Goldstein, W.H., Osterheld, A., Oreg, J., \& Bar-Shalom, A. 1989, 
ApJ, 344, L37

\bibitem[]{}
Kastner, S.O., \& Kastner, R.E. 1990, JQSRT, 44, 275

\bibitem[]{}
Mariska, J.T., Feldman, U., \& Doschek, G.A. 1979, A\&A, 73, 361

\bibitem[]{}
Mazzotta, P., Mazzitelli, G., Colafrancesco, S., \& Vittorio, N. 1998,
A\&AS, 133, 403

\bibitem[]{}
Meyer, J.P., 1985, ApJS, 57, 173

\bibitem[]{}
Rosner, R., Tucker, W.H., \& Vaiana, G.S. 1978, ApJ, 220, 643

\bibitem[]{}
Phillips, K.J.H., Greer, C.J., Bhatia, A.K., Coffey, I.H., Barnsley, R., 
\& Keenan, F.P. 1997, A\&A, 324, 381

\bibitem[]{}
Rugge, H.R., \& McKenzie, D.L. 1985, ApJ, 297, 338

\bibitem[]{}
Saba, J.L.R., Schmelz, J.T., Bhatia, A.K., \& Strong, K.T. 1999, ApJ, 
510, 1064

\bibitem[]{}
Schmelz, J.T., Saba, J.L.R., Chauvin, J.C., \& Strong, K.T. 1997, ApJ, 
477, 509

\bibitem[]{}
Schrijver, C.J., \& McMullen, R.A. 2000, ApJ, in press

\bibitem[]{}
Smith, B.W., Mann, J.B., Cowan, R.D., \& Raymond, J.C.1985, ApJ, 298, 898

\bibitem[]{}
Waljeski, K. et al. 1994, ApJ, 429, 909

\bibitem[]{}
Witt, A.N., 1977, ApJS, 35, 1

\bibitem[]{}
Wood, K., \& Reynolds, R.J. 1999, ApJ, 525, 799

\bibitem[]{}
Yusef-Zadeh, F., Morris, M., \& White, R.L. 1984, ApJ, 278, 186



\end{thebibliography}
\end{document}